\documentclass[aps,prb,reprint,superscriptaddress]{revtex4-1}
\usepackage{amsmath}
\usepackage{amssymb}
\usepackage{graphicx}
\usepackage{color,xcolor}

\begin{document}

\title{Noncollinear ferrielectricity and morphotropic phase boundary in GeS monolayer}

\author{Shi-Xin Song}
\affiliation{School of Physics,
             Southeast University,
             Nanjing 211189, China}

\author{Yang Zhang}
\affiliation{Department of Physics and Astronomy,
             University of Tennessee,
             Knoxville, Tennessee 37996, USA}

\author{Jie Guan}
\email{guanjie@seu.edu.cn}%
\affiliation{School of Physics,
             Southeast University,
             Nanjing 211189, China}

\author{Shuai Dong}
\email{sdong@seu.edu.cn}%
\affiliation{School of Physics,
             Southeast University,
             Nanjing 211189, China}

\date{\today}

\begin{abstract}
Two-dimensional polarity is intriguing but remains in the early stage. Here a structural evolution diagram is established for GeS monolayer, which leads a noncollinear ferrielectric $\delta$-phase energetically as stable as the ferroelectric $\alpha$-phase. Its ferrielectricity is induced by the phonon frustration, i.e., the competition between ferroelectric and antiferroelectric modes, providing more routes to tune its polarity. Besides its prominent properties like large band gap, large polarization, and high Curie temperature, more interestingly, the morphotropic phase boundary between $\alpha$- and $\delta$-phases is highly possible, which is crucial to obtain giant piezoelectricity for lead-free applications.
\end{abstract}

\maketitle

\section{Introduction}
Ferroelectric (FE) materials with spontaneous switchable electric polarization have been widely explored, for their rich physics and broad applications \cite{Rabe:mp,Scott:sci,Ahn:Rmp}. However, for those mostly used FE perovskite oxides, the ferroelectricity is seriously suppressed when the material thickness decreases to $\sim1$ nm, due to the strong depolarization field and surface effects \cite{thickness1,thickness3}, which limits the further miniaturization of devices.

Two-dimensional (2D) van der Waals (vdW) polar materials provide the possibility to overcome this challenge and achieve atomic-scale applications \cite{Wu:WIRCMS,Kou:JPCL}. Recently, many 2D FE materials, e.g., SnTe~\cite{SnTe}, CnInP$_2$S$_6$~\cite{CIPS,You:sa}, SnSe~\cite{Chang:Science, Chang:PRL}, and SnS~\cite{Higashitarumizu:nc}, have been obtained in experiments. More 2D FE materials were theoretically predicted \cite{Ding:nc,Fei:prl,Huang:prl,Xiang:Prl}. In addition, antiferroelectricity with antiparallel electric dipoles was also found in MXene \cite{Maxene17} and group V monolayers~\cite{Xiao18}. Besides these plain collinear dipole textures, noncollinear dipole order was also recently predicted in monolayers of dioxydihalides (WO$_2$Cl$_2$ and MoO$_2$Br$_2$) \cite{WOCl}. These progresses have opened an emerging field of 2D polar materials.

Despite these achievements, the physical understanding of 2D polarity remains in the early stage, comparing with the three-dimensional (3D) counterpart. For example, the noncollinear dipole orders have been experimentally realized in perovskite deviants \cite{Ramesh,Belik}, but the experimental verification in aforementioned 2D dioxydihalides remains challenging since they are fragile against moisture. Thus it is crucial to find other 2D robust systems to host noncollinear dipole orders and verify the physical mechanism. Another vital concept in 3D FE bulks is the so-called morphotropic phase boundary (MPB) between two similar phases with different symmetries \cite{ShiranePRB}. The MPB's in Pb(Zr$_x$Ti$_{1-x}$)O$_3$ (PZT) and Pb(Mg$_{0.33}$Nb$_{0.67}$)O$_3$-PbTiO$_3$ (PMN-PT) lead to giant piezoelectricity, and thus have been commercially used in microelectronics, optoelectronics, and microelectromechanical systems \cite{ShiranePRL,YE2008}. Thus, the discovery of MPB in other polar systems, e.g. BiFeO$_3$ \cite{Zeches977}, is highly desired for lead-free piezoelectric applications. However, the MPB has not been reported in 2D polar systems yet.

In this Letter, we study a 2D noncollinear ferrielectric (FiE) system $\delta$-GeS, with prominent polar properties. In fact, the 2D group IV-VI family, including GeS, GeSe, SnS, etc, forms a rich mine of 2D polar materials \cite{aGeS,SnSe2,SnSNL, barrazalopez:RMP}. Moreover, multiple stable phases have been found due to their pyramid-like $3$-fold bonding~\cite{bGeS,SnSe1,GeSe17,GeSe19,Brownson:CM}, similar to phosphorene~\cite{DT232}, which provides the possibility to host exotic dipole textures. Indeed, here the phase coexisting of $\delta$-GeS and $\alpha$-GeS is proposed, which leads to the MPB in the 2D limit. Our work is based on density functional theory (DFT), and details of methods can be found in Supplementary Materials (SM) \cite{GeS20SM}.

\begin{figure}
\includegraphics[width=0.48\textwidth]{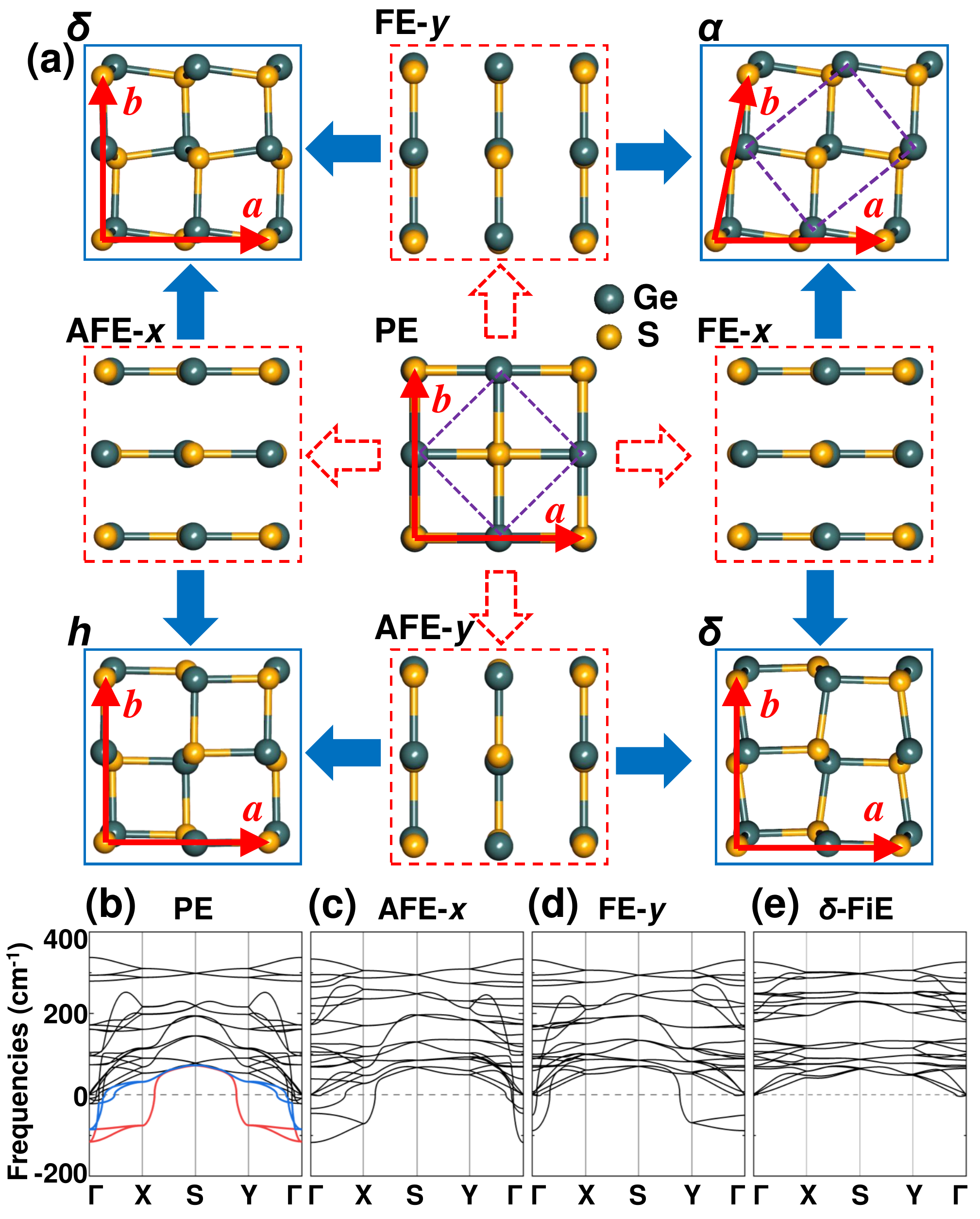}
\caption{(a) Structural evolution diagram of GeS monolayer. Center: the square PE structure as parent phase. The dynamic stable phases are four corner ones, which can be derived by condensing two branches of unstable phonon modes together. The edge ones, obtained by considering only one single branch of unstable phonon mode for each other, are dynamic unstable. For comparison among different phases, the base vectors $a-b$ are shown, while the minimal unit cells of PE and $\alpha$-phase are indicated by green rectangles. (b-e) The corresponding phonon spectra.}
\label{fig1}
\end{figure}

\section{Results \& Discussion}
\subsection{Origin of $\delta$-phase.}
In Ref.~\cite{aGeS}, the so-called $\alpha$-GeS (space group No. 31 $Pnm2_1$) was predicted, together with other $\alpha$-phases of group IV-VI family. The $\alpha$ phase can be derived from the paraelectric (PE) parent phase (space group No. 129 $P4/nmm$), as shown in Fig.~\ref{fig1}(a).

Using a $\sqrt{2}\times\sqrt{2}$ unit cell [constructed by the $a$-$b$ vectors in Fig.~\ref{fig1}(a)], the phonon spectrum of PE phase has two (dual) main imaginary frequencies at the Brillouin center $\Gamma$ point, as shown in Fig.~\ref{fig1}(b). The duality comes from the tetragonal symmetry. As a result, a uniform in-plane displacement of all Ge atoms can occur along the $<110>$ direction (i.e. the $<100>$ direction of the primitive cell), leading to the FE $\alpha$-phase as predicted in Ref.~\cite{aGeS}, whose phonon spectrum shows its dynamic stability despite of tiny imaginary frequencies at $\Gamma$ point, as shown in Fig. S1(d) in SM \cite{GeS20SM}.

However, beside the imaginary frequencies at the $\Gamma$ point, there are additional imaginary frequencies at the Brillouin edge $X/Y$ points, which indicate the AFE distortions. Considering the $x$-$y$ duality from tetragonal symmetry, the FE and AFE orders can be along the $x$ or $y$ axis. However, the pure FE or AFE order along the $x$ or $y$ axis remains dynamic unstable, with residual imaginary frequencies [Fig.~\ref{fig1}(c-d)]. Only the combination of two distortion modes can lead to stable phases. For example, the $\alpha$-phase can be considered as a combination of FE modes along the $x$ and $y$ axes. In this sense, the combination of FE mode along the $x$ axis and AFE mode along the $y$ axis (or vice versa) can lead to a new phase, i.e., the $\delta$-GeS (space group No. 29 $Pca2_1$) \footnote{Ref.~\cite{Ma:Nanotechnology} once studied this structure of GeS but did not touched its structural evolution and polarity.}, which owns a noncollinear dipole texture, similar to the case of dioxydihalides \cite{WOCl}. Such $\delta$-GeS shares the identical geometry of $\delta$-phosphorene \cite{DT232}. The phonon spectrum of $\delta$-GeS [Fig.~\ref{fig1}(e)] indicates its dynamic stability. Furthermore, the combination between two AFE modes along the $x$ and $y$ axes can lead to one more exotic noncollinear dipole texture, i.e., the so-called Haeckelite phase ($h$-GeS), which is also dynamic stable. The complete diagram of phase evolution is sketched in Fig.~\ref{fig1}(a).

To further verify the stability of $\delta$-GeS monolayer, the cohesive energy is calculated, to compare with previously reported $\alpha$- and $\beta$-GeS \cite{aGeS,bGeS}. As summarized in Table~\ref{Table1}, the cohesive energy of $\delta$-GeS is very close to that of $\alpha$-GeS, both of which are higher than those of $\beta$-phase and $h$-phase. Since the $\alpha$-GeS monolayer/few-layer had been experimentally realized \cite{GeS17NR,GeS18CM}, the predicted $\delta$-GeS should be also available, considering the proximate energy (only $2.1$ meV/atom higher) \footnote{Ref.~\cite{Brownson:CM} reported a new phase of GeS bulk, which was named as $\delta$-GeS. However, the concrete structural information was not revealed except the lattice constants.}. Since the energy of $h$-phase is higher ($12.9$ meV/atom higher than $\alpha$-phase), in the following only the $\delta$-phase will be focused on.

\begin{table}
\centering
\caption{Comparison among different phases of 2D GeS monolayers. $a$ and $b$ are the in-plane lattice constants as defined in Fig.~\ref{fig1}(a). $E_{\rm coh}$ is the cohesive energy per atom with respect to isolated atoms. $P$ is the polarization. Our results of $P$ for $\alpha$-GeS and $\beta$-GeS monolayers are in consistent with previous works \cite{aGeS,bGeS}. The $\beta$-phase is not derived from the aforementioned square structure \cite{bGeS}.}
\begin{tabular*}{0.48\textwidth}{@{\extracolsep{\fill}}llllll}
\hline
\hline
         &  PE & $\alpha$  & $\delta$  & $h$ & $\beta$\\
\hline
$a$ (\AA)   & $5.517$ & $5.778$ & $5.754$ & $5.657$ & $3.492$ \\
$b$ (\AA)   & $5.517$ & $5.778$ & $5.680$ & $5.656$ & $5.648$\\
$E_{\rm coh}$ (eV/atom) &  $3.5976$   & $3.6216$ & $3.6195$  & $3.6087$ & $3.6030$ \\
$P$ (pC/cm) & $0$ & $4.81$  & $2.73$  & $0$ & $1.95$ \\
\hline
\hline
\end{tabular*}
\label{Table1}
\end{table}

\begin{figure}
\includegraphics[width=0.48\textwidth]{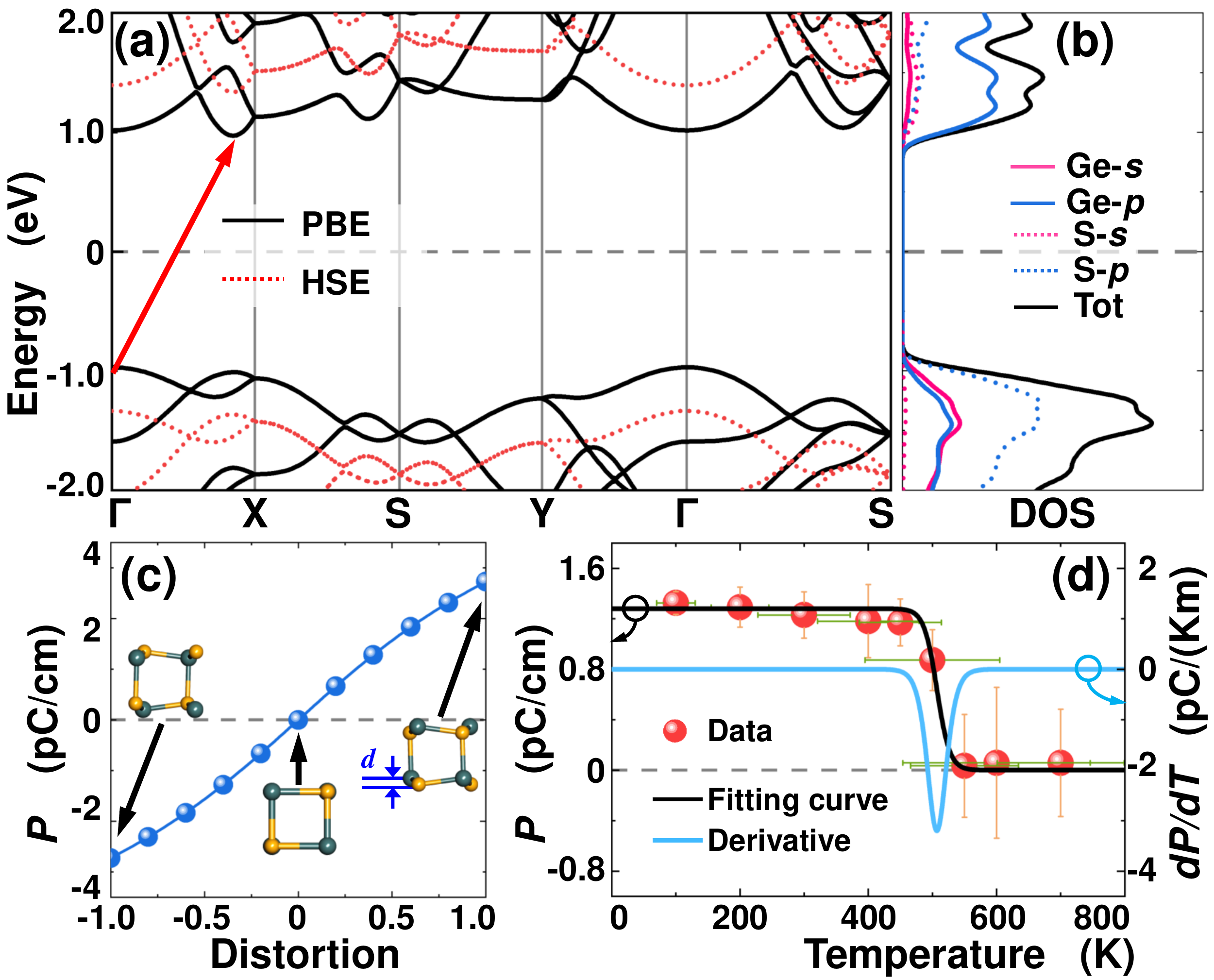}
\caption{(a) Electronic band structure (in both PBE and HSE levels) of $\delta$-GeS monolayer. An indirect band gap is indicated by the red arrow. (b) The corresponding PDOS. (c) Distortion dependent polarization $P$ of $\delta$-GeS mononlayer. Insert: the distortion is characterized by the atomic displacement $d$, normalized to its optimized one. (d) Left axis: $P$ as a function of temperature $T$ from AIMD simulation. Right axis: The corresponding pyroelectric response $dP/dT$.}
\label{fig2}
\end{figure}

The electronic band structure of $\delta$-GeS monolayer is shown in Fig.~\ref{fig2}(a). The $\delta$-GeS monolayer is an insulator, with an indirect band gap ($1.94$ eV in PBE and $2.67$ eV in HSE06), whereas the shapes of band dispersion are similar between PBE and HSE06. Such an appropriate band gap of $\delta$-GeS allows the switching of its polarization by external electric field.

The projected density of states (PDOS) is plotted in Fig.~\ref{fig2}(b). There is significant hybridization between the Ge's $s$-/$p$-orbitals and S's $p$-orbitals at the valence band maximum (VBM), while the conduction band minimum (CBM) is mostly contributed by Ge's $p$-orbitals. The Bader charge analysis suggests that $0.84$ electron is transferred from Ge to S \cite{Bader06}. It is the interplay of both ionic and covalent bonding between Ge and S atoms that causes the lattice distortions in $\delta$-GeS~\cite{Kooi221}, which is different from the driving force in WO$_2$Cl$_2$ and MoO$_2$Br$_2$ (the so-called $d^0$ rule) \cite{WOCl}. Such ionic and covalent bonding requires the structure of GeS satisfies the 3-fold bonding, i.e., each Ge bonds with three neighboring S and vice versa. All aforementioned stable phases fulfill this rule.

By taking the S framework as the reference, Ge ions in the other sub-layer have the zigzag-type displacements in the $\delta$-phase. The displacements ($d$) along the $y$-axis [insert of Fig.~\ref{fig2}(c)] are uniform, leading to a net FE polarization. Meanwhile, the displacements along the $x$-axis have opposite directions between neighboring Ge-S chains along the $x$-axis, i.e., with the AFE dipole order and thus zero net polarization. Using the standard Berry phase calculations \cite{Vanderbilt:prb}, the FiE polarization (along the $y$-axis) is estimated as $2.73$ pC/cm [Fig.~\ref{fig2}(c)] for $\delta$-GeS, which is lower than that of FE $\alpha$-phase (as compared in Table~\ref{Table1}). If an effective thickness of $\delta$-GeS monolayer can be evaluated as the half of lattice constant $c$ in the AB stacking bulk structure (i.e., $5.37$ {\AA} \cite{GeS20SM}), the 3D polarization is estimated as $50.8$ $\mu$C/cm$^2$, about twice of traditional FE perovskite BaTiO$_3$ ($20\sim25$ $\mu$C/cm$^2$)~\cite{BTOpolarization2}.

More interestingly, such ferrielectricity is robust against thermal fluctuation. We performed the {\it ab initio} molecular dynamics (AIMD) simulation for a $\delta$-GeS monolayer to estimate its FiE transition temperature ($T_{\rm C}$). The net polarization of $\delta$-GeS was calculated by the point charge model as a function of temperature, as shown Fig.~\ref{fig2}(d). A distinct phase transition occurs around $500$ K. This $T_{\rm C}$ makes $\delta$-GeS monolayer attractive for room-temperature applications.

\begin{figure}
\includegraphics[width=0.48\textwidth]{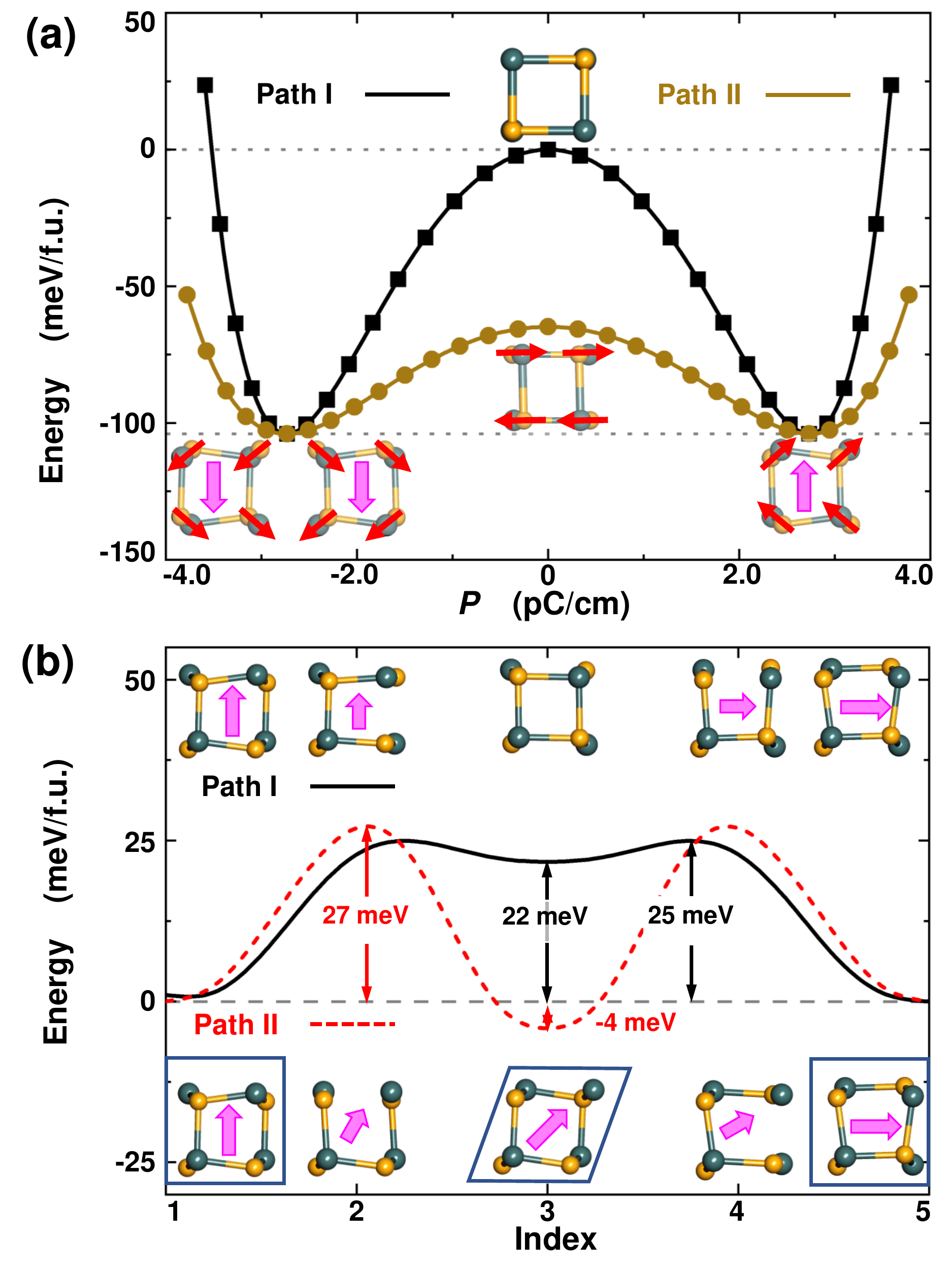}
\caption{(a) Double-well energy profiles for two FiE switching processes. The path I corresponds to the $180^\circ$ in-plane flipping of dipoles, while the easier path II corresponds to the $83^\circ$ in-plane rotation of dipoles. (b) The possible two paths for $90^\circ$ rotation of FiE domain obtained using the nudged elastic band (NEB) method \cite{NEB2}}
\label{fig3}
\end{figure}

\subsection{Domain switching \& morphotropic phase boundary.}
In the following, the physics beyond a static FiE single domain will be studied, which is nontrivially interesting for $\delta$-GeS monolayer.

First, for any polar material, the polarization switching is a key physical process. For the noncollinear FiE system, the switching paths are more interesting than those plain FE cases. As shown in Fig.~\ref{fig3}(a), the $180^\circ$ switching of net polarization can be achieved via two paths. The path II, with $83^\circ$ in-plane rotation of local dipoles, can be easier than the path I with direct $180^\circ$ in-plane flip of local dipoles. This is a unique characteristic of noncollinear FiE system.

Second, the $90^\circ$ rotation of net polarization, i.e., the change between twin domains, can be realized via two paths, as shown in Fig.~\ref{fig3}(b). The path I involves the $h$-phase as the intermediate one, while the path II involves the $\alpha$-phase. The energy barriers are close between these two paths ($27$ meV/f.u. \textit{vs} $25$ meV/f.u.), but the path II owns a much lower energy saddle point.

\begin{figure}
\includegraphics[width=0.48\textwidth]{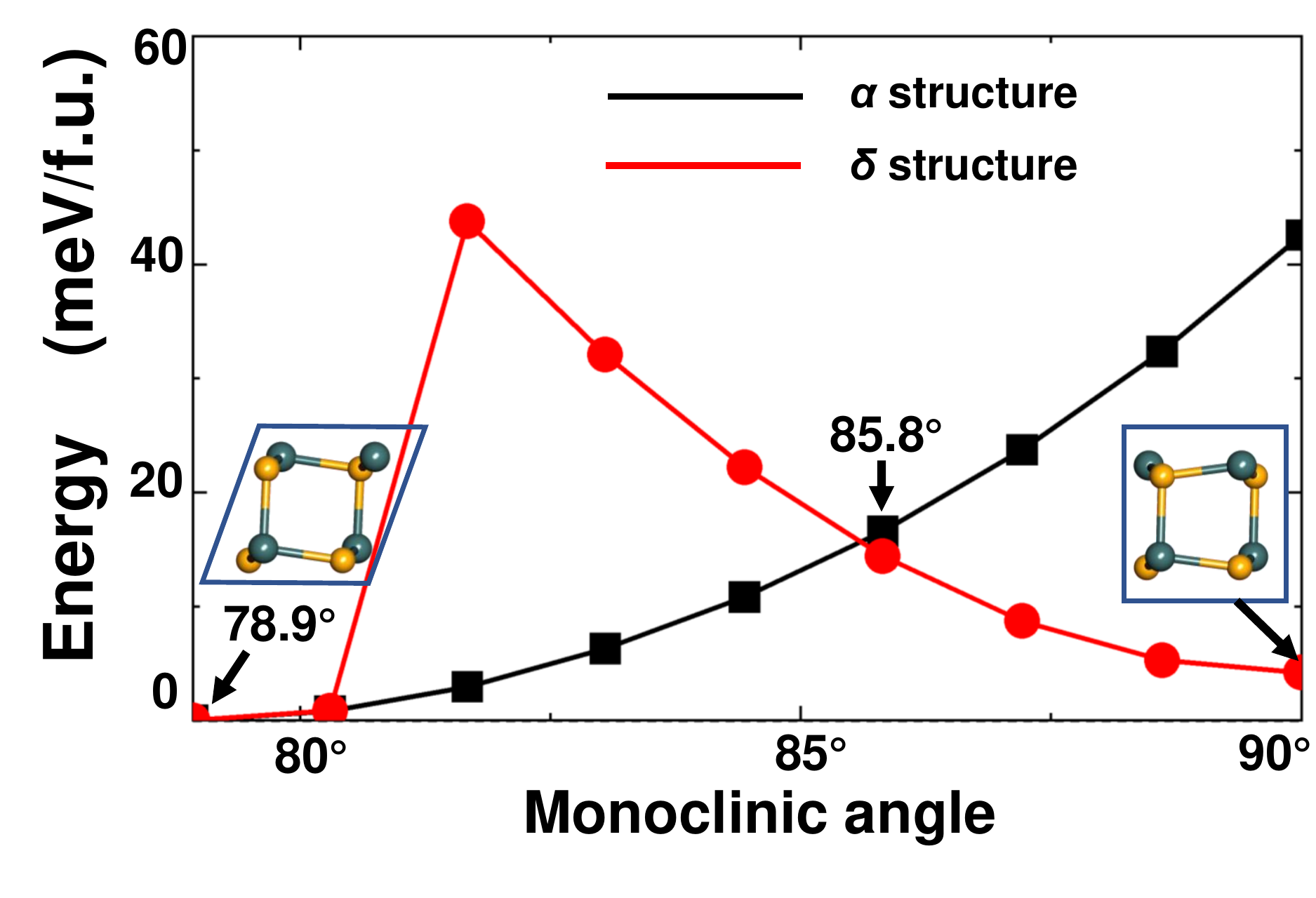}
\caption{Strain modulated phase transition between the $\alpha$- and $\delta$-phases of GeS monolayer. By continuously tuning the monoclinic angle between vectors $a$-$b$ ($90^\circ$ for the $\delta$-phase and $78.9^\circ$ for the $\alpha$-phase), the energies of $\alpha$- and $\delta$-phase can be reversed. The intermediate framework are obtained using the linear interpolation between the optimized $\alpha$- and $\delta$-phases. When the monoclinic distortion is large enough, the $\delta$-phase decays to the $\alpha$-phase spontaneously.}
\label{fig4}
\end{figure}

Since the energies of $\alpha$-GeS and $\delta$-GeS are very proximate, their stability should be sensitive to stimulus, and thus tunable. As shown in Fig.~\ref{fig4}, the energy of $\alpha$-GeS will increase if the monoclinic distortion is suppressed. Then the $\delta$-phases will be the stabler one if the lattice is close to tetragonal, which can be realized by certain shearing strain conditions. The proximate energies and lattice constants, but different symmetries and polarizations, can lead to MPB between $\alpha$- and $\delta$-phases of GeS monolayer, which will be the origin of giant piezoelectric effect and many exotic collective effects.

\begin{figure}
\includegraphics[width=0.48\textwidth]{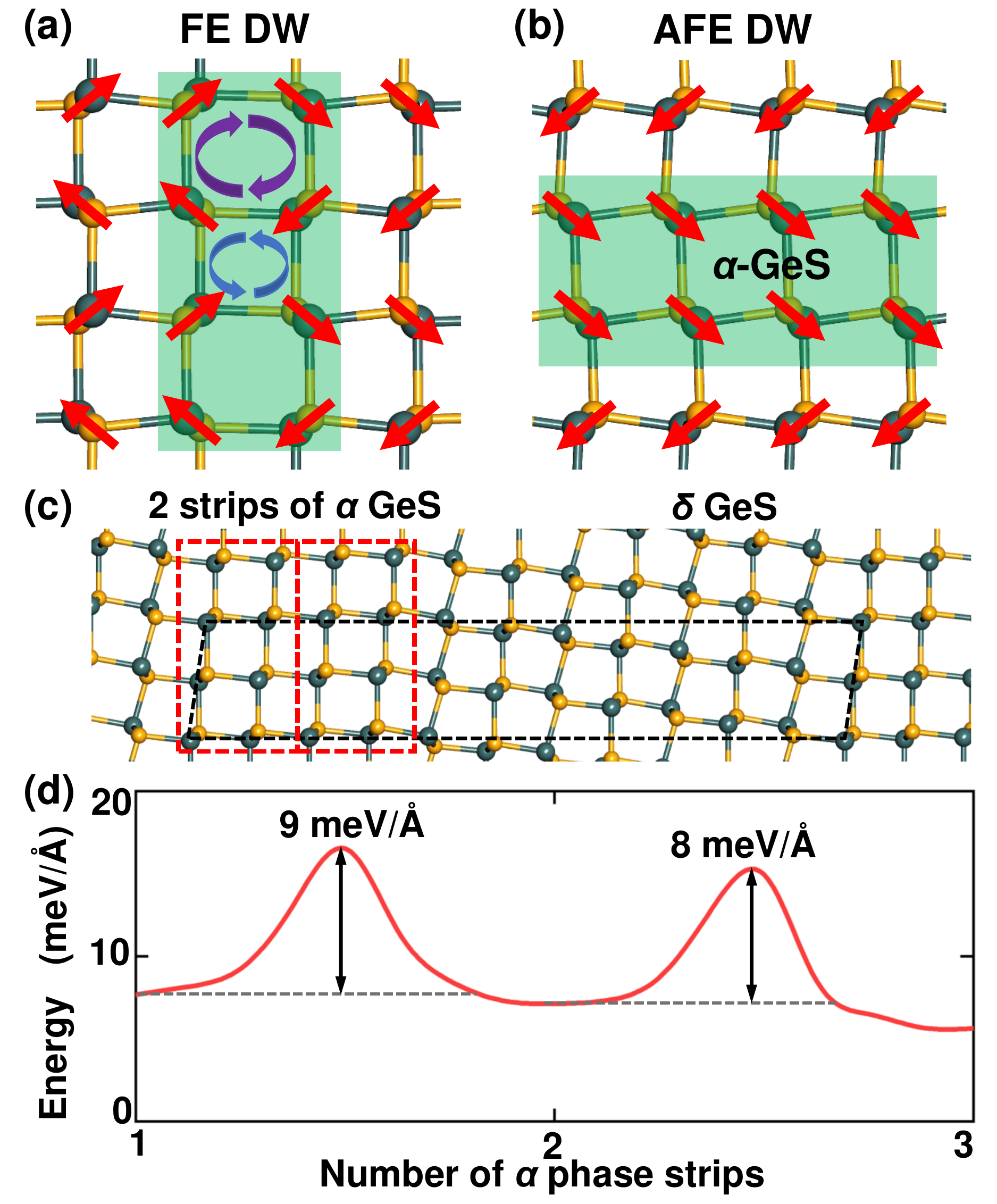}
\caption{Phase coexistence in GeS monolayer. (a) The FE domain wall of $\delta$-GeS is a strip of $h$-phase. (b) The AFE domain wall of $\delta$-GeS is a strip of $\alpha$-phase. Such domain walls, i.e., the phase coexistence, are stable during the structural relaxation. (c) Schematic of the coexistence of $\alpha$- and $\delta$-phases, with two strips of $\alpha$-phase for example. (d) The width of $\alpha$-phase can change with energy barriers.}
\label{fig5}
\end{figure}

The strain-dependent MPB will lead to possible phase coexistence, which is essential physics in many strongly correlated electronic systems \cite{Dagotto257} as well as relaxor ferroelectrics \cite{Li2016ncomm}. In fact, the domain walls in $\delta$-GeS can be a seed of phase coexistence. As shown in Fig.~\ref{fig5}(a-b), the FE domain wall propagating along the $b$-axis is just the $h$-phase, while the AFE domain walls propagating along the $a$-axis is just the $\alpha$-phase. Both these domain walls are atomically sharp and stable during the structural relaxation. Since these two types of domain walls does not break the local 3-fold bonding character, the domain wall energies are not high: $26$ meV/{\AA} for the FE wall and $2$ meV/{\AA} for the AFE one, respectively.

The phase coexistence can occur in even larger scale. To demonstrate this issue, here a large supercell is constructed, with partial $\alpha$ region and partial $\delta$ region, as shown in Fig.~\ref{fig5}(c). The stability of phase boundary is verified by structural relaxation. By deducting the energies of $\alpha$ region and $\delta$ region, the energy for phase boundary can be estimated, which is about $3.2\pm 1.0$ meV/{\AA}. By changing the size of $\delta$ region, the energy barriers for phase boundary shift is also estimated, as shown in Fig.~\ref{fig5}(d). Such energy barriers make the MPB stable in GeS monolayer, which will not decompose to a single phase spontaneously.

As mentioned before, according to the experience in 3D piezoelectric crystals and the related physical principles, it is well accepted that the MPB will enhance the piezoelectricity. Although the DFT method can not directly calculate the piezoelectricity of MPB system here, other numerical methods in larger scales, like the phase field model simulation, are encouraged to verify the enhanced piezoelectricity in near future.

\section{Conclusion}
In summary, a frustration-induced noncollinear dipole order has been revealed in the ferrielectric $\delta$-GeS monolayer, which is as stable as the ferroelectric $\alpha$-GeS monolayer. Its prominent ferrielectric properties are attractive for experimental verifications and room-temperature applications. Furthermore, the competition between $\alpha$-phase and $\delta$-phase can be tuned by strain, and they can coexist with the morphotropic phase boundary. Our results can also be extended to other IV-VI two-dimensional polar systems, e.g., $\delta$-SnSe \cite{deltaSnSe}, which will significantly add value to low-dimensional materials.

\begin{acknowledgments}
We thank Jun Chen for useful discussions. Work were supported by the National Natural Science Foundation of China (Grant Nos. 61704110 and 11834002), the Fundamental Research Fund for the Central Universities, the Shuangchuang Doctoral Program of the Jiangsu Province, and the Zhongying Young Scholar Program of Southeast University. We thank the Big Data Computing Center of Southeast University for providing
facility support for performing calculations.
\end{acknowledgments}

\bibliographystyle{apsrev4-1}
\bibliography{ges20ref2}
\end{document}